\documentclass[twocolumn,showpacs,preprintnumbers,amsmath,amssymb]{revtex4}

\usepackage{graphicx}
\usepackage{dcolumn}
\usepackage{bm}

\begin{document}

\title{Magnetic properties of a spin-3 Chromium condensate}
\author{Liang He and Su Yi}
\affiliation{Key Laboratory of Frontiers in Theoretical Physics, Institute of
Theoretical Physics, Chinese Academy of Sciences, Beijing 100190, China}

\begin{abstract}
We study the ground state properties of a spin-3 Cr condensate subject to an
external magnetic field by numerically solving the Gross-Piteavskii equations.
We show that the widely adopted single-mode approximation is invalid under a
finite magnetic field. In particular, a phase separation like behavior may be
induced by the magnetic field. We also point out the possible origin of the
phase separation phenomenon.
\end{abstract}

\date{\today}

\maketitle

\section{Introduction}

Since the realization of Bose-Einstein condensate of chromium
atoms~\cite{Pfau05}, there have been considerable experimental and theoretical
efforts in exploring physical properties of chromium condensates. Owing to the
large magnetic dipole moment of chromium atoms, the dipolar effects was first
identified experimentally from its expansion dynamics~\cite{stuh}. More
remarkably, with the precise control of the short-range interaction using
Feshbach resonance, the $d$-wave collapse of a pure dipolar condensate has been
observed~\cite{laha}.

In the context of spinor condensates, chromium atom has an electronic spin
$s=3$, which provides an ideal platform for exploring even richer quantum
phases as compared to those offered by the spin-1 and spin-2
atoms~\cite{Ho98,Ohmi98,Law98,Ciobanu00,Koashi00,Stenger1998,Barret01,
Schmaljohann04,Chang04, Kuwamoto04}. To date, theorists have mapped out the
detailed phase diagram of a spin-3 chromium
condensate~\cite{Santos06,Ho06,Makela07}. In particular, a more exotic biaxial
nematic phase was also predicted~\cite{Ho06}. The possible quantum phases and
defects of spin-3 condensates were also classified based on the symmetry
considerations~\cite{barnett,yip}. Other work on spin-3 chromium condensates
includes theoretically studying the strongly correlated states of spin-3 bosons
in optical lattices~\cite{bernier} and the Einstein-de Haas effect in chromium
condensates~\cite{Santos06,Ueda}.

Nevertheless, all the previous work concerning the ground state and the
magnetic properties of spin-3 Cr condensates has adopted the so-called
single-mode approximation (SMA), which assumes that all spin components share a
common density profile. However, the studies on spin-1 case show that, for an
antiferromagnetic spinor condensate, SMA is invalid in the presence of magnetic
field for antiferromagnetic spin exchange interaction~\cite{Yi02(SMA)}. One
would naturally question the validity of SMA for spin-3 condensate since the
short-range interactions involved here are more complicated than those in
spin-1 system.

In the present paper, we study the ground state properties of a spin-3 chromium
condensate subject to a uniform axial magnetic field by numerically solving the
Gross-Pitaevskii equations. We show that even though SMA is still valid in the
absence of an external magnetic field, it fails when the magnetic field is
switched on. More remarkably, we find that when the undetermined scattering
length corresponding to total spin zero channel falls into a certain region,
the magnetic field may induce a phase separation like behavior such that the
peak densities of certain spin components do not occur at the center of the
trapping potential.

This paper is organized as follows. In Sec. \ref{form}, we introduce our model
for numerical calculation. The results for the ground state structure of a
spin-3 condensate under an external magnetic field are presented in Sec.
\ref{resu}. Finally, we conclude in Sec. \ref{con}.

\section{Formulation}\label{form}
We consider a condensate of $N$ spin $s=3$ chromium atoms subject to a uniform
magnetic field ${\mathbf{B}}=B{\mathbf{z}}$. In mean-field treatment, the
system is described by the condensate wave functions $\psi_m$
($m=-3,-2,\ldots,3$). The total energy functional of the system,
$E[\psi_m,\psi_m^*]$, can be decomposed into two parts $E=E_0+E_1$ with $E_0$
and $E_1$ being, respectively, the single-body and interaction energies.
Adopting the summation convention over repeated indices, the single-body energy
can be expressed as
\begin{eqnarray}
E_0\!=\!\int \!d{\mathbf{r}}\psi_m^*\left[\left(-\frac{\hbar^2\nabla^2}{2M}+V_{\mathrm{ext}
}\right)\delta_{mm^{\prime}}+g\mu_BBs^z_{mm^{\prime}}\right]\psi_{m^{\prime}}, \nonumber\\
\end{eqnarray}
where $M$ is the mass of the atom, the trapping potential
$V_{\mathrm{ext}}({\mathbf r})=\frac{1}{2}M\omega_\perp^2(x^2+y^2+\eta^2z^2)$
is assumed to be axially symmetric with $\eta$ being the trap aspect ratio,
${\mathbf{s}}=(s^x,s^y,s^z)$ are the spin-3 matrices, $g=2$ is the Land\'{e}
$g$-factor of $^{52}$Cr atoms, and $\mu_B$ is Bohr magneton.

\begin{figure*}[tbp]
\centering
\includegraphics[width=6.8in]{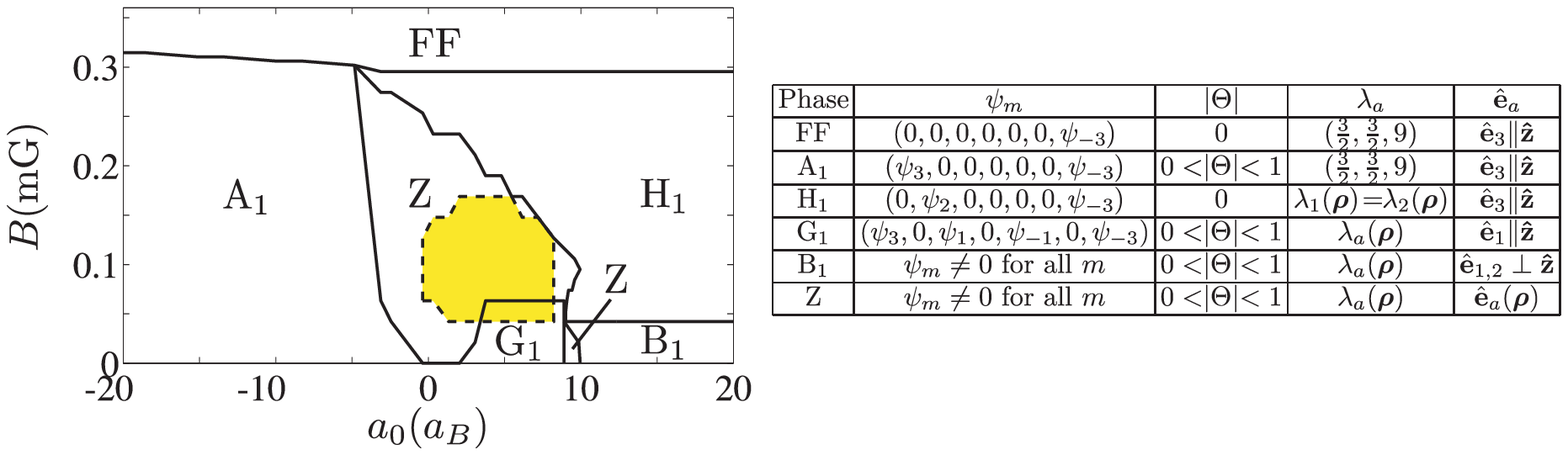}
\caption{(Color online) Left panel: phase diagram of spin-3 Cr condensate in
the $a_0$-$B$ parameter space. The shaded region indicates the region where
phase separation occurs (see text for details). Right panel: the main
characteristics of the quantum phases.} \label{phases}
\end{figure*}

The collisional interaction between two spin-3 atoms takes the
form~\cite{Ho98,Ohmi98}
\begin{eqnarray}
V_{\mathrm{int}}(\mathbf{r},\mathbf{r}^{\prime })=\delta (\mathbf{r}-\mathbf{r}
^{\prime })\sum_{S=0}^{2s}g_{S}\mathcal{P}_{S},\label{inter}
\end{eqnarray}
where $\mathcal{P}_S$ projects onto the state with total spin $S$ and
$g_S=4\pi\hbar^2a_S/M$ with $a_{S=0,2,4,6}$ being the scattering lengths for
the combined symmetric channel $S$. For $^{52}$Cr, it was determined
experimentally that $a_{6}=112\,a_{B}$, $a_{4}=58\,a_{B}$, and
$a_{2}=-7\,a_{B}$ with $a_B$ being Bohr radius~\cite{stuh}, while the value of
$a_{0}$ is unknown, and we shall treat it as a free parameter in the results
presented below. Making use of the relations~\cite{Ho98}
\begin{eqnarray}
1&=&\sum_S\mathcal{P}_S,\nonumber\\
{\mathbf{s}}_1\cdot{\mathbf{s}}_2&=&\sum_S\frac{\mathcal{P}_S}{2}[S(S+1)-24], \nonumber\\
({\mathbf{s}}_1\cdot{\mathbf{s}}_2)^2&=&\sum_S\frac{\mathcal{P}_S}{4} [S(S+1)-24]^2,\nonumber
\end{eqnarray}
we may replace $\mathcal{P}_2$, $\mathcal{P}_4$, and $\mathcal{P}_6$ in Eq.
(\ref{inter}) by $1$, ${\mathbf{s}}_1\cdot{\mathbf{s}}_2$, and
$({\mathbf{s}}_1\cdot{\mathbf{s}}_2)^2$, such that the interaction energy
functional becomes
\begin{eqnarray}
E_{1}=\frac{1}{2}\int d{\mathbf{r}}n^2\left[C+\alpha|\Theta|^2+\beta{\rm Tr}{\mathcal N}^2+
\gamma\langle{\mathbf s}\rangle^2\right],  \label{E1}
\end{eqnarray}
where the interaction parameters are
$C=-\frac{1}{7}g_{4}+\frac{81}{77}g_4+\frac{1}{11}7g_{6}$,
$7\alpha=g_0-\frac{5}{3}g_2+\frac{9}{11}g_4-\frac{5}{33}g_6$, $\beta
=\frac{1}{126}g_{2}-\frac{1}{77}g_4+\frac{1}{198}g_{6}$, and
$\gamma=-\frac{5}{84}g_2+\frac{1}{154}g_4+\frac{7}{132}g_6$. Furthermore,
$n({\mathbf{r}})=\psi_m^*\psi_m$ is the total density, $$\Theta({\mathbf{r
}})=\frac{1}{n}\sqrt{7}\langle
00|3m;3m^{\prime}\rangle\psi_m\psi_{m^{\prime}}$$ is the singlet amplitude, and
$$\langle{\mathbf{s}}\rangle({\mathbf
r})=\frac{1}{n}\psi_m^*{\mathbf{s}}_{mm^{\prime}}\psi_{m^{\prime}}$$ is the
density of spin. Finally,
$$\mathcal{N}_{ij}({\mathbf{r}})=\frac{1}{2n}
\psi^*_m(s^is^j+s^js^i)_{mm^{\prime}}\psi_{m^{\prime}},\quad i,j=x,y,z$$ is the
nematic tensor, and to obtain it, we have utilized the relation
$$\langle ({\mathbf s}_1\cdot{\mathbf s}_2)^2\rangle={\rm Tr}{\cal N}^2-\frac{1}{2}\langle{\mathbf s}_1\rangle\cdot\langle {\mathbf s}_2\rangle.$$
The nematic tensor was first introduced in the liquid crystal physics as the
order parameter $\mathcal{N}$~\cite{de Gennes} to describe the orientation
order of the liquid crystal molecules. Since $\mathcal{N}$ is Hermitian, it can
be diagonalized with all eigenvalues $\lambda_{a=1,2,3}$ (ordered as
$\lambda_1\leq\lambda_2\leq\lambda_3$) being real and the corresponding
principle axes $\hat{\mathbf{e}}_{a}$ being mutually orthogonal. Unless all
three eigenvalues are equal, the systems with two identical eigenvalues are
usually refer to as uniaxial nematics, while those with three unequal
eigenvalues are biaxial ones. More importantly, $\lambda_a$ can be determined
by performing Stern-Gerlach experiments along $\hat{\mathbf{e}}_{a}$
\cite{Ho06}. As it can be seen from Eq. (\ref{E1}), different quantum phases
originate from the competition of $\Theta$, $\langle\mathbf{s}\rangle$, and
$\mathcal{N}$. Following the discussion of Diener and Ho~\cite{Ho06}, we shall
characterize phases in a spin-3 Cr condensate using the condensate wave
functions $\psi_m$, singlet amplitude $\Theta$, spin
$\langle{\mathbf{s}}\rangle$, and nematic tensor $\mathcal{N}$.

\begin{figure*}[tbp]
\centering
\includegraphics[width=6.9in]{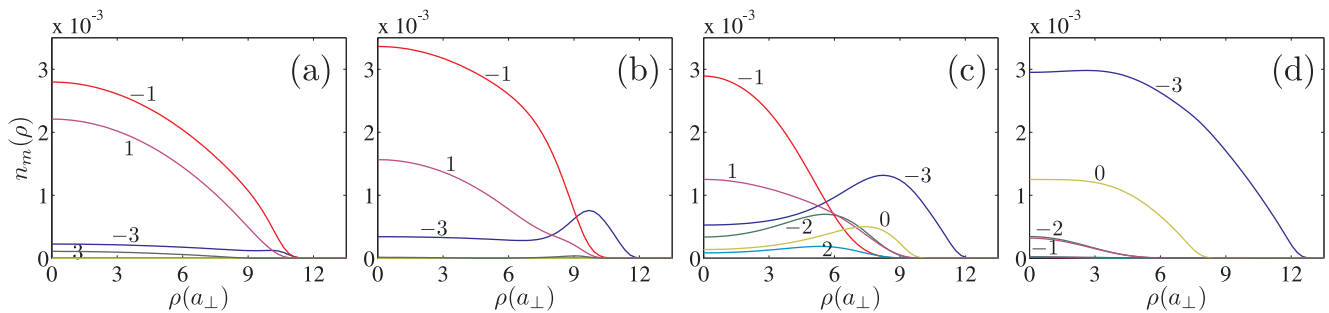}
\caption{(Color online) The typical $\protect\rho$ dependence of the densities
for all spin components in the phase separation region. From (a) to (d), the
magnetic field strengths (in units of mG) are, respectively, $B=0.0211$,
$0.0633$, $0.0844$, and $0.1689$. The scattering length $a_0=5.47a_B$ is the
same for all figures. The densities of those components not shown in the
figures are too small to be seen.} \label{phsep}
\end{figure*}

To simplify the numerical calculations, we shall focus on highly oblate trap
geometries ($\eta\gg1$) such that the condensate can be regarded as
quasi-two-dimensional whose motion along the $z$-axis is frozen to the ground
state of the axial harmonic oscillator. The condensate wave functions can then
be decomposed into
\begin{eqnarray}
\psi_m({\mathbf{r}})=(\eta/\pi)^{1/4}e^{-\eta z^2/2}\phi_m({\boldsymbol{\rho}})
\end{eqnarray}
with ${\boldsymbol{\rho}}=(x,y)$ and $\phi_m$ being normalized to the total
number of atoms $N$, i.e., $\int d{\mathbf{r}}\phi_m^*\phi_m=N$. After
integrating out the $z$ variable, $E_0$ gives an extra constant, while the
interaction parameters $C$, $\alpha$, $\beta$, and $\gamma$ are all rescaled by
a factor $(\eta/2\pi)^{1/2}$. The mean-field wave functions $\{\psi_m\}$ are
obtained by minimizing the total energy functional numerically using imaginary
time evolution. We shall focus our study on the Cr line \cite{Ho06}, namely
only the scattering length $a_0$ is allowed to changed freely, since
experimentally, it is the most relevant case. For all results presented in the
present work, we have chosen $N=10^5$, $\omega_\perp=2\pi\times
100\,\mathrm{Hz}$, and $\eta=10$. Correspondingly, the dimensionless length
unit $a_\perp=\sqrt{\hbar/(M\omega_\perp)}$ is adopted in throughout this
paper.

We remark that we have neglected the magnetic dipole-dipole interaction energy
in Eq. (\ref{E1}) for simplicity, as in the present work, we are concentrating
on investigating how short-range interaction and magnetic field affect the
ground state wave function. The ignorance of dipolar interaction in spinor Cr
condensate was also justified in Ref. \cite{Ho06}. Moreover, we have
numerically confirmed that for the parameter used in this paper, the
dipole-dipole interaction energy is much smaller than short-ranged
spin-dependent interaction energy when the condensate is not polarized by the
magnetic field.

\section{Results}\label{resu}
Figure \ref{phases} summarizes the main results of this paper. In the left
panel of Fig. \ref{phases}, we present the phase diagram of spin-3 Cr
condensate in the $a_0$-$B$ parameter space, here we have adopted the similar
notations for different phases as in Ref. \cite{Ho06}. In the right panel, we
tabulate the major characters of each phase. We remark that the resolution of
phase diagram is limited by step sizes of $a_0$ and $B$ when we numerically
scan the parameter plane, therefore, it is possible that more details may
emerge by reducing the step sizes.

\subsection{Condensate wave functions}
The numerical results indicate that the condensate wave functions can always be
expressed as
\begin{eqnarray}
\phi_m({\boldsymbol{\rho}})=\sqrt{n_m({\rho})}e^{i\vartheta_m},
\end{eqnarray}
where the density of $m$th component $n_m({\rho})$ is an axially symmetric
function and the corresponding phase $\vartheta_m$ is a constant independent of
the spatial coordinates. In case the external magnetic field is completely
switched off, we find that all wave functions $\phi_m({\mathbf r})$ have the
same density profile, indicating that the SMA is valid for spin-3 condensates
in the absence of magnetic field. We note that this conclusion also holds true
for spin-$1$ and -$2$ condensates.

Once the external magnetic field is applied, SMA quickly becomes invalid. More
remarkably, as shown in Fig. \ref{phsep}, when control parameters $a_0$ and $B$
fall into the shaded region in the left panel of Fig. \ref{phases}, the peak
densities of at least one of the spin components among $m=-3$, $\pm 2$, and $0$
do not occur at the center of the trap, in analogy to the phase separation in a
two-component condensate~\cite{binexp,binary}. In the absence of magnetic
field, the system is symmetric under SO(3) rotation of the spin, and $m=-3$
component can be populated. However, immediately after we switch on the
magnetic field, this SO(3) symmetry is broken such that $m=\pm 1$ spin
components are highly populated under a very weak magnetic field. As one
continuously increases the magnetic field, the occupation number in $m=-3$
component increases with density in the margin of the trap growing faster than
that in the center, which induces the phase separation like behavior. When the
population in $m=-3$ component dominates, the peak densities of all spin
component occur at the center of the trap. We remark that similar behavior of
the wave functions also appears outside the Cr line~\cite{Ho06}.

\begin{figure}[tbp]
\includegraphics[width=3.in]{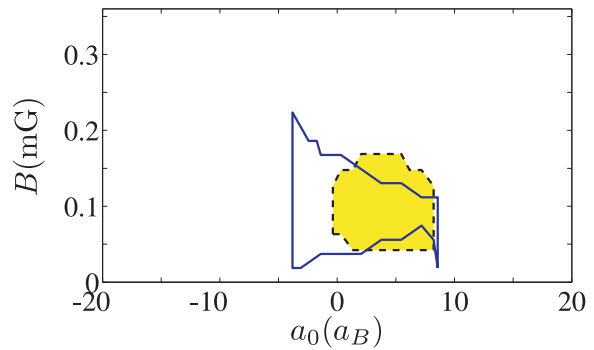}
\caption{(color online) The instable region (enclosed by solid line) of a homogeneous Cr condensate and the phase separation region (enclosed by dashed line) of a trapped Cr condensate (Same as that in Fig. \ref{phases}).}
\label{instab}
\end{figure}

To gain more insight into the origin of the phase separation like behavior, we
consider a homogeneous Cr condensate where each spin component has already
condensed into the zero momentum mode. The wave function $\psi_m({\mathbf r})$
for phase unseparated state is then replaced by a uniform $c$-number
\begin{eqnarray}
\bar\psi_m=\sqrt{n}\xi_m,
\end{eqnarray}
where $n$ is a real constant and $\xi_m$ are complex constants. The ground
state can be obtained by minimizing the total energy $E$ subject to the
normalization condition $\xi_m^*\xi_m=1$. In such a way, we have reproduce the
phase diagram in Ref. \cite{Ho06}. To confirm that those phases are indeed the
ground states, we introduce a new set of variables, $\zeta_p$ and
$\zeta_{p+1}$, corresponding to, respectively, the real and imaginary parts of
the wave function $\xi_m$ as
\begin{eqnarray}
\zeta_{p=2(3-m)+1}={\rm Re}[\xi_m] \mbox{ and } \zeta_{p=2(3-m)+2}={\rm Im}[\xi_m].\nonumber
\end{eqnarray}
We then construct the Hessian matrix ${\boldsymbol H}=\left[\frac{\partial^2 E}
{\partial \zeta_p\partial \zeta_q}\right]$. For a solution to be stable, the
Hessian matrix must be positive definite~\cite{timmer}. In Fig. \ref{instab},
we present the unstable region of a homogeneous Cr condensate on $a_0$-$B$
plane. To obtain it, we have chosen the density to be $n=3.3\times
10^{14}\,{\rm cm}^{-3}$ which is the peak density of the trapped system in our
numerical calculations. One immediately sees that the unstable region of a
homogeneous condensate roughly agrees with the phase separation region of the
trapped system, which suggests that the possible origin of the phase separation
behavior is the instability of the phase unseparated solution.

We emphasize that, unlike in a binary Bose-Einstein condensate where the
emergence of phase separation is determined by the strengths of intra- and
inter-species interactions, here for a given scattering length $a_0$, the phase
separation like behavior is induced by the magnetic field.

\subsection{Singlet amplitude}
\begin{figure}[tbp]
\centering
\includegraphics[width=2.8in]{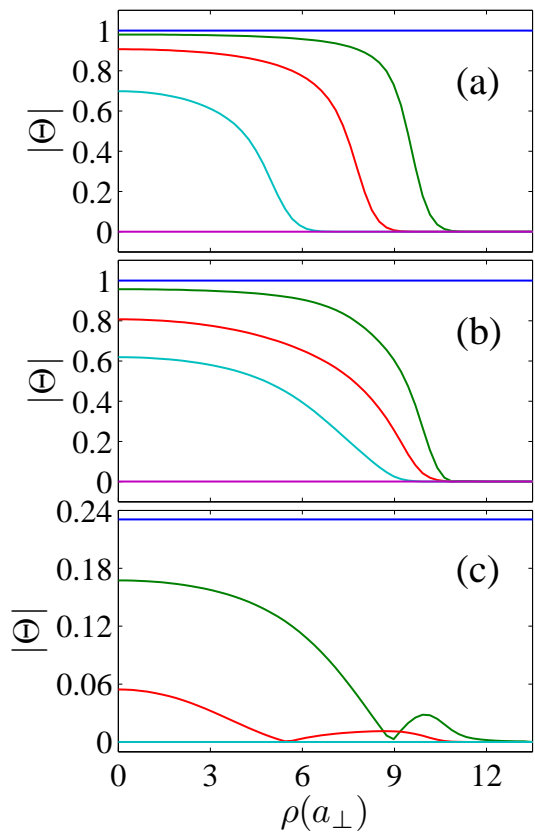}
\caption{(Color online) The typical behaviors of $|\Theta(\protect\rho)|$ for $a_0=-8.27a_B$ (a), $5.47a_B$ (b), and $12.35a_B$ (c). In descending order of central value, the lines in (a) correspond to the magnetic field (in units of mG) $B=0$, $0.0244$, $0.1689$, $0.2533$, and $0.3377$; those in (b) correspond to $B=0$, $0.0422$, $0.0844$, $0.1266$, and $0.19$; and finally, those in (c) correspond to $B=0$, $0.0211$, $0.0422$, and $0.0633$.}
\label{theta}
\end{figure}

Since the spatial independence of $\Theta({\mathbf{\rho}})$ is a necessary
condition for SMA, it can also be used as a criterion to check the validity of
SMA. As shown in Fig. \ref{theta}, $|\Theta|$ is a constant when $B=0$; while
immediately after the magnetic field is turned on, $|\Theta|$ becomes spatially
dependent. In addition, the peak value of $|\Theta|$ decreases continuously as
one increases the magnetic field until it completely vanish.

In Fig.~\ref{theta}~(a), $|\Theta(\rho)|$ becomes zero only after the
condensate is completely polarized, while in (b) and (c), it vanishes once the
system enters the ${\rm H}_1$ phase. Therefore, using singlet amplitude, we may
map out the phase boundaries between ${\rm A}_1$ and ${\rm FF}$, ${\rm Z}$ and
${\rm H}_1$, and ${\rm B}_1$ and ${\rm H}_1$. However, $\Theta$ alone is
incapable of determining other phase boundaries. Finally, we note that, for
$a_0>8.9a_B$, the value of $|\Theta|$ drops much faster with the increasing
magnetic field than that corresponding to $a_0<8.9a_B$, as shown below this
behavior has a direct impact on the magnetization curve of the system.

\begin{figure}[tbp]
\centering
\includegraphics[width=3.2in]{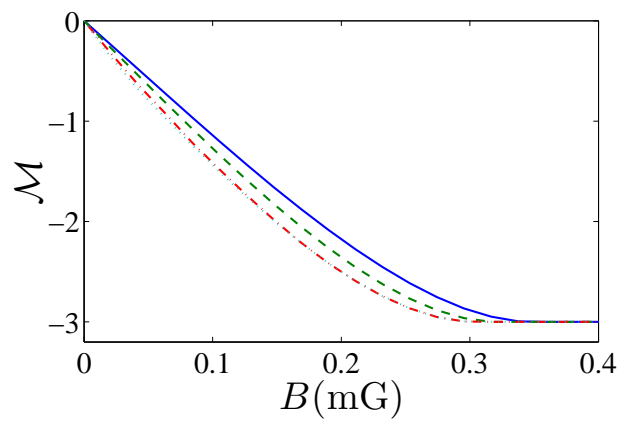}
\caption{(Color online) The field dependence of the magnetization for $a_0=-59.82a_B$ (solid line), $-25.45a_B$ (dashed line), $5.47a_B$ (dash-dotted line), and $12.35a_B$ (dotted line). For $a_0>8.9a_B$ ($\alpha>0$), the $\mathcal{M}(p)$ curves corresponding to different $a_0$'s are indistinguishable.}
\label{mag}
\end{figure}

\subsection{Magnetization}
We now turn to study magnetic field dependence of the total magnetization. To
this end, we define the reduced magnetization as
\begin{eqnarray}
\mathcal{M}=N^{-1}\int d{\mathbf{\rho}}\langle s^z\rangle.\label{magm}
\end{eqnarray}
Unlike in Ref.~\cite{Makela07} where the total magnetization is conserved, here
we allow it to change freely. Therefore, the transverse components of the spin,
$\langle s^x\rangle$ and $\langle s^y\rangle$, are always zero. Figure
\ref{mag} shows the field dependence of the reduced magnetization, which
approaches $-3$ when $B$ reaches the saturation field. We note that, for
$a_0<8.9a_B$, the behavior of $\mathcal{M}(B)$ slightly depends on the value of
$a_0$; while for $a_0>8.9a_B$, the magnetization curves corresponding to
different $a_0$'s become indistinguishable. Consequently, as shown in Fig.
\ref{phases}, the saturation field for the former case is a decreasing function
of $a_0$, while for the latter one, it becomes a constant. The $a_0$
independence of the magnetization for $a_0>8.9a_B$ case can be qualitatively
understood as follows. The scattering length $a_0$ only contributes to the
total energy through singlet amplitude $\Theta$. As shown in Fig. \ref{theta},
for $a_0>8.9a_B$, $\Theta$ vanishes quickly as one increases the magnetic
field, such that varying $a_0$ only yields a negligible effect on magnetization
curve. With the help magnetization, we can further identify the phase boundary
between ${\rm H_1}$ and ${\rm FF}$ phases.

\begin{figure}[tbp]
\centering
\includegraphics[width=3.3in]{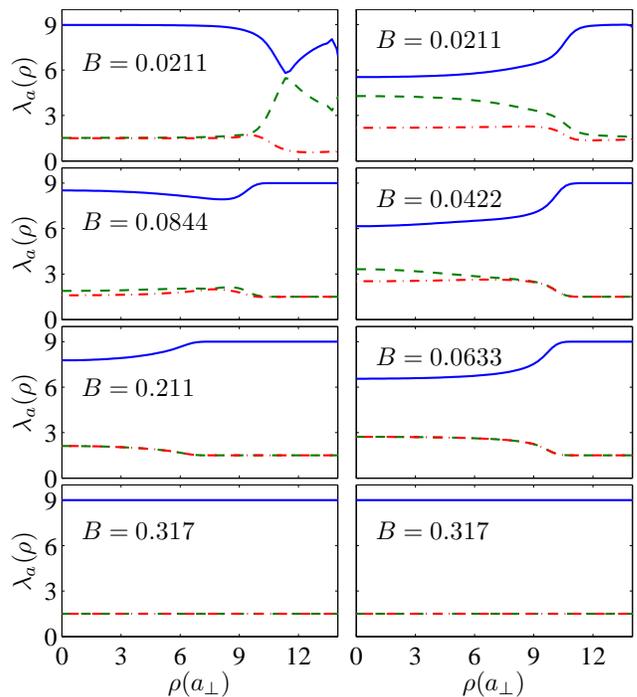}
\caption{(Color online) The spatial dependence of $\lambda_1$ (dash-dotted lines), $\lambda_2$ (dashed lines), and $\lambda_3$ (solid lines) for $a_0=5.47a_B$ (left panels) and $12.35a_B$ (right panels). The magnetic field strength (in units of mG) is denoted in each figure.}
\label{nematic}
\end{figure}

\subsection{Nematic tensor}
To determine other phase boundaries, we have to rely on the nematic tensor. In
Fig. \ref{nematic}, we plot typical behavior of the eigenvalues of nematic
tensor under different $a_0$ and magnetic fields. As a consequence of the
failure of SMA, $\lambda_a$'s are generally spatially dependent. However, some
of their characteristics obtained under SMA remain to be true.

The full ferromagnet (FF) phase occurs when the magnetic field exceeds the
saturation field such that only $m=-3$ component is occupied. The nematic
tensor takes a diagonal form with $\lambda_1=\lambda_2=\frac{3}{2}$ and
$\lambda_3=9$. For the wave functions of A$_1$ and H$_1$ phases, only two spin
states are populated: other than a common $m=-3$ state, $m=3$ and $2$ are also
occupied for, respectively, A$_1$ and H$_1$ phases. One can easily deduce that
the nematic tensors of phases A$_1$ and H$_1$ are, respectively, ${\rm
diag}\{\frac{3}{2},\frac{3}{2},9\}$ and $n^{-1}{\rm
diag}\{\frac{3}{2}n_{-3}+4n_2,\frac{3}{2}n_{-3}+4n_2, 9n_{-3}+4n_2\}$. Since
for both cases, ${\cal N}_{zz}$ is the largest eigenvalue, we have
$\hat{\mathbf e}_3\parallel \hat{\mathbf z}$. Moreover, $\hat{\mathbf e}_1$ and
$\hat{\mathbf e}_2$ are spatially independent.

For G$_1$ phase, $m=\pm2$ and $0$ states are unoccupied, consequently, ${\cal
N}$ becomes a block diagonal matrix with one of the eigenvalues being ${\cal
N}_{zz}=n^{-1}\left[9(n_{3}+n_{-3})+n_{1}+n_{-1}\right]$. Furthermore, from
numerical results, we find that ${\cal N}_{zz}$ is always the smaller
eigenvalue and $\lambda_2({\boldsymbol\rho})\neq\lambda_3({\boldsymbol\rho})$,
which suggests that G$_1$ is a biaxial nematic phase with $\hat {\mathbf
e}_1\parallel\hat {\mathbf z}$.

All spin components of ${\rm B}_1$ and ${\rm Z}$ phases are populated, they are
all are biaxial nematic with three unequal and spatially dependent $\lambda$'s.
The principal axes for these phases are also spatially dependents. The only
difference is that for ${\rm B}_1$ phase we can identify that either ${\mathbf
e}_1$ or ${\mathbf e}_2$ is perpendicular to $z$-axis.

\section{Conclusions}\label{con}
To conclude, we have mapped out the phase diagram of a spin-3 Cr condensate
subject to an external magnetic field based on the numerical calculation of the
ground state wave function. In particular, we show that SMA becomes invalid for
Cr condensates under a finite magnetic field. More remarkably, if the unknown
scattering length $a_0$ falls into the region $[-0.37,8.2]a_B$, a phase
separation like behavior may be induced by the magnetic field. We also point
out that such behavior might originate from the instability of a phase
unseparated solution. As a future work, we shall investigate the ground state
structure of a spin-3 Cr condensate by including the magnetic dipole-dipole
interaction.

\acknowledgments
We thank Han Pu for useful discussion. This work is supported by NSFC (Grant No. 10674141), National 973 program (Grant No. 2006CB921205), and the ``Bairen" program of the Chinese Academy of Sciences.

\end{document}